\documentclass[english]{article}
\usepackage[T1]{fontenc}
\usepackage[latin1]{inputenc}
\usepackage{amsmath}
\usepackage{graphicx}
\usepackage{amssymb}

\makeatletter

\newcommand{\noun}[1]{\textsc{#1}}
\providecommand{\tabularnewline}{\\}

 \newcommand{\lyxaddress}[1]{
   \par {\raggedright #1 
   \vspace{1.4em}
   \noindent\par}
 }

\date{}
\usepackage{ae,aecompl}
\usepackage[colorlinks=true, linkcolor=blue, 
                   citecolor=blue, urlcolor=blue]{hyperref}

\usepackage{babel}
\makeatother
\begin{document}

\title{The Effect of Clouds on Air Showers Observation from Space }

\author{T. Abu-Zayyad$^{\textrm{1 *}}$, C.C.H. Jui$^{\textrm{1}}$, E.C.
Loh$^{\textrm{1}}$}

\maketitle

\lyxaddress{\begin{center}\emph{\small $^{\textrm{1}}$University of Utah, Department
of Physics and High Energy Astrophysics Institute, Salt Lake City,
Utah, USA}\end{center}}

\begin{abstract}
Issues relating to extensive air showers observation by a space-borne
fluorescence detector and the effects of clouds on the observations
are investigated using Monte Carlo simulation. The simulations assume
the presence of clouds with varying altitudes and optical depths.
Simulated events are reconstructed assuming a cloud-free atmosphere.
While it is anticipated that auxiliary instruments, such as LIDAR
(LIght Detection And Ranging), will be employed to measure the atmospheric
conditions during actual observation, it is still possible that these
instruments may fail to recognize the presence of a cloud in a particular
shower observation. The purpose of this study is to investigate the
effects on the reconstructed shower parameters in such cases. Reconstruction
results are shown for both monocular and stereo detectors and for
the two limiting cases of optically thin, and optically thick clouds. 
\end{abstract}
\newpage

\section{\label{sec:Introduction}Introduction}

Space-borne cosmic rays detectors for energies E $\geq10^{20}$ eV
have been proposed \cite{Takahashi_95} and are now under study \cite{red_book}.
Such a detector will comprise one or two satellites orbiting the Earth
at an altitude of $\sim$~400~km to $\sim$~1000~km and will have
a wide field of view (FOV), on the order of 60$^{o}$. The footprint
on the Earth's surface of the FOV has dimensions on the same order
of magnitude as the orbit height. Studies of the global distribution
of clouds and their frequency of occurrence, e.g. \cite{Wylie_94},
suggest that the target volume will at any point in time contain some
clouds.

The amount of clouds (fractional cover), the distribution of clouds
in terms of cloud type, altitude, and optical depth will undoubtedly
affect the detector's trigger aperture. In addition, cloud presence
could result in a reduction of the reconstructible aperture, as contaminated
events are excluded from the analysis. Finally cloud presence could
compromise the accuracy of the energy estimate for an observed event,
since this estimate depends in part on a knowledge of the atmospheric
conditions at the time and location of the shower development and
along the path the light from the shower travels to the detector.

The effect of cloud presence on the detector aperture is beyond the
scope of this paper. In this study we limit our attention to the question
of how cloud presence may affect the reconstructed shower geometry
and energy. In the context of a Monte Carlo study, this question can
be addressed by applying the event analysis assuming no cloud presence,
and then determine (a) whether the reconstruction procedure can identify
the presence of otherwise unreported clouds, and thereby rejecting
the event in question, and (b) for those events where clouds eluded
all detection attempts, how the reconstructed shower parameters were
altered.

With respect to the detector itself, there are two possible modes
of operation: monocular and stereo, the latter employing two sites
(satellites) separated by some distance and which view the same region
of the sky. The Fly's Eye experiment has demonstrated the superiority
of the stereo technique on the ground \cite{Cassiday_90}. For space-borne
detectors, it has been suggested \cite{EUSO_proposal} that monocular
observation can perform as well as stereo if use is made of the information
provided by the reflection of the \v{C}erenkov beam associated with
the shower off the surface of the Earth, in order to reconstruct the
shower geometry. In this study we also investigate possible errors
introduced in cases where the reflection occurs off the top of a cloud
instead of the surface.

The answer to (a) above will depend on whether or not cloud presence
will manifest itself through a significant alteration in the expected
detector response to the shower signal. As an example, the reflectivity
of an optically opaque cloud is several times larger than that of
the surface of the ocean, $\sim80$\%-90\% vs. $\sim10$\%-20\%, therefore
a test may be be developed and applied to an individual shower observation
looking at the signal strength of the last few pixels to determine
whether the reflection of the beam has occurred off the top of a cloud.
The development of such a test is not trivial. It must be applicable
to a wide range of shower energies and geometries as well as accommodate
different atmospheric conditions and cloud optical properties. Also,
the formulation of such a test must rely on a detailed description
of the event data recorded by the detector for each shower observation.
As described in section \ref{sub:The-Detector}, we do not attempt
a detailed simulation of the detector data acquisition system and
event formation logic. Also, we only treat a few combinations of shower
geometries and clouds configurations. Therefore, the development of
a test for cloud presence based on the event data is beyond the scope
of this study.

Clouds come in a wide variety of cloud types, heights, vertical extent,
and optical depths. There are, in general, also spatially in-homogeneous
and finite, a few kilometers in lateral extent, clouds. This makes
a general treatment of all possible scenarios difficult. To simplify
the discussion we will concentrate on two limiting cases. The first
case is that of a high altitude, optically thin cloud. This case corresponds
to cirrus clouds which are pervasive in the atmosphere \cite{Wylie_94}.
The second case is that of low altitude, optically thick clouds. These
types of clouds are easy to detect in general but may be difficult
to detect under some circumstances, e.g., if the cloud is small in
lateral size (on the order of a few kilometers.) 

This paper is organized as follows: The next section provides a motivation
for the different cloud configurations used in the study. Following
that is an overview of the Monte Carlo simulation of the detector,
showers, and the atmosphere including cloud simulation. Section \ref{sec:Event-reconstruction}
provides an overview of the shower geometry and energy reconstruction
procedures. Finally section \ref{sec:Results} presents the results
of the study.

\section{\label{sec:Clouds-and-EAS}Clouds and EAS}

Clouds are classified as (a) low-level (cloud base height, $h_{base}<2$~km),
(b) mid-level ($2<h_{base}<6$~km), and (c) high-level( $h_{base}>6$~km)
\cite{Liou_ch4}. In the equatorial region, high-level clouds (Cirrus)
typically occur at altitudes of $8-15$~km \cite{Liou_ch4}. Most
EAS develop in the lower atmosphere at altitudes~<~20~km, where
most of the atmospheric mass is located. Depending on their altitude,
clouds are made up of predominantly water molecules, (low-level),
mixed water molecules and ice crystals (mid-level), and ice crystals
(high-level). For our purposes, water and ice crystal clouds have
one important difference: the number density of scatterers. The concentration
of water molecules in a low level cloud is a factor of 10 to 100 greater
than that of ice crystals in a high altitude cirrus cloud. This results
in a much smaller scattering length for the low level cloud, and the
relation between the optical depth of the cloud and its physical thickness
becomes qualitatively different. 

From detailed Monte Carlo simulations, most EAS generated by protons
or nuclei in the energy range $10^{19}\leq E\leq10^{21}$~eV reach
maximum development at atmospheric slant-depths, $x$, between 700-1000~gm/cm$^{2}$
(depending on the energy, primary type, and the hadronic model used
in the simulations) \cite{Pryke_2k1}. Beyond the shower maximum depth,
$x_{max}$, the number of electrons in the shower falls rapidly. 

\begin{figure}
\includegraphics[%
  scale=0.5]{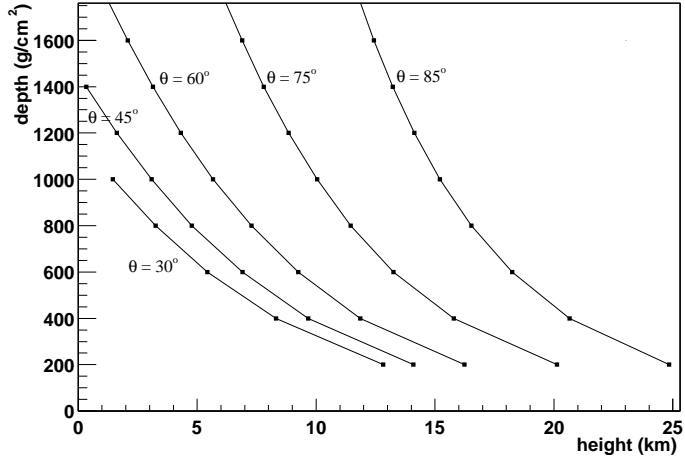}

\caption{\label{fig:slant_vs_height} Slant depth (g/cm$^{2}$) along a shower
track with a zenith angle, $\theta$, vs. the height (km) above the
surface of a point along the shower track. Data from table \ref{table:h_vs_slant}.}
\end{figure}

To quantify how clouds might affect space-borne observations of extensive
air showers, we need to relate the atmospheric slant depth along the
shower track to altitude above the Earth's surface (sea level). This
relation depends on the shower zenith angle and is presented in figure
\ref{fig:slant_vs_height} and table \ref{table:slant_vs_h}. As can
be seen from the table, different cloud altitude and shower zenith
angle combinations can result in the shower front reaching the cloud
top at different stages of the shower development. Three broad cases
can be identified: (1) clouds above shower development, (2) clouds
in the region of shower development, and \emph{}(3) clouds below shower
development.

\begin{table}

\caption{\label{table:h_vs_slant} The height (km) above the surface, of a
point along the shower track at a given slant depth (gm/cm$^{2})$
along the shower, for showers with different zenith angles, $\theta$.
Atmospheric density profile according to the U.S. Standard Atmosphere,
1976.\vspace{1ex}}

\begin{tabular}{|c|r|r|r|r|r|}
\hline 
$x_{slant}$ \textbackslash{} $\theta$ &
30&
45&
60&
75&
85\tabularnewline
\hline 
200&
12.81&
14.08&
16.23&
20.13&
24.84\tabularnewline
\hline 
400&
8.31&
9.67&
11.85&
15.79&
20.65\tabularnewline
\hline 
600&
5.43&
6.89&
9.25&
13.25&
18.24\tabularnewline
\hline 
800&
3.24&
4.79&
7.28&
11.45&
16.53\tabularnewline
\hline 
1000&
1.46&
3.07&
5.68&
10.04&
15.21\tabularnewline
\hline
1200&
-&
1.62&
4.32&
8.85&
14.13\tabularnewline
\hline
1400&
-&
0.34&
3.14&
7.80&
13.22\tabularnewline
\hline 
1600&
-&
-&
2.07&
6.88&
12.43\tabularnewline
\hline
\end{tabular}
\end{table}

In addition to the location of the cloud in relation to the shower-detector
geometry, the optical depth, $\tau$, and physical thickness, $\Delta z$,
of the cloud also play a role in determining what the detector sees.
Hence they also need to be considered in combination with the height
of the cloud top, $h_{top}$. Finally, depending on the values of
$\tau$ and $\Delta z$, the effects of multiple light scattering
may or may not be negligible, $\lesssim$10\%. 

As mentioned in the introduction, we will only address the two limiting
cases for the optical depth. The reason for this is that the consideration
of these cases is sufficient for the study of the relevant problems
of: light transmission through high altitude cirrus clouds (optically
thin), and the \v{C}erenkov beam reflection off optically thick clouds
and its effect on the monocular geometry reconstruction. In general,
single scattering calculations are sufficient for a medium $\tau<0.1$.
First order corrections may be required for $0.1<\tau<0.3$ \cite{van_de_Hulst,Liou_2k_laser_trans}.
From our own studies, we saw that for $\tau<0.5$, second order corrections
account for less than $\sim$~10\% of the total signal transmitted
through a cloud. Therefore, to avoid having to calculate light multiple
scattering beyond first order corrections we will restrict our definition
of optically-thin clouds to mean clouds with optical depth $\tau<0.5$. 

For the case of an opaque cloud, a light beam impinging on the top
of the cloud will be reflected as a result of a large number of multiple
scatterings inside the cloud. The amount, spatial and temporal distributions,
and the direction of the reflected photons can be calculated using
a Monte Carlo procedure (see section \ref{sub:The-case-tau-gt3}).
The use of this procedure or an equivalent detailed simulation of
the cloud reflective properties would be required if one is to attempt
to infer the cloud presence from the event data. For this study, however,
it is sufficient to treat reflection off clouds in the same fashion
as reflection off the surface, as described in section \ref{sub:Cerenkov-Spot-sec_Sim}.

\section{\label{sec:Simulation}Simulation}

Both the simulation and reconstruction programs used in this study
are based on those developed for the High Resolution Fly's Eye (HiRes)
experiment \cite{AbuZ_2k_nima}. The original programs are described
in detail in \cite{AbuZ_Thesis}. While the underlying algorithms
are similar to those used by HiRes, the actual code was converted
from Fortran and C to the C++ language and extensive use was made
of the ROOT data analysis framework \cite{ROOT_NIMA}. 

Naturally, the HiRes detector simulation was replaced by a description
of the OWL detector. Otherwise, a large portion of the simulation
code, e.g. the atmosphere, is detector independent, and was retained.
Cloud simulation including the effects of multiple light scattering
was added, and minor modifications were made throughout the code to
account for the differences between the two detectors.

In the following subsections a description of the detector simulation
is given, followed by a synopsis of the atmospheric modeling. We then
present an overview of the shower simulation, and the section ends
with a discussion of the \v{C}erenkov spot.

\subsection{\label{sub:The-Detector}The Detector}

A description of the OWL baseline instrument is given in \cite{owl_roadmap}.
Earlier versions of proposed designs were presented in a workshop
\cite{red_book}. Our detector simulation is based on these earlier
designs. The conclusions drawn from these studies will not be substantially
affected by the design evolution. A description of the simulated detector
follows:

The detector consists of two cameras, each mounted on a satellite.
The satellites have an orbital height of 800~km and are separated
by a distance of 500~km. Each camera comprises a concave, spherical
mirror and a focal plane detector. The mirror has an effective light
collecting area of 4.9~m$^{2}$ and a field of view (FOV) corresponding
to a cone of half-angle of 30$^{o}$. The mirrors axes are tilted
slightly from the nadir in order for the two cameras to view a common
area on the surface.

For the reason that the optical design of the detector had not been
completed at the time this study was begun, and also for the sake
of simplicity, we opted to use a scheme in which all photon ray-tracing
calculations are done in angular space. In this case we ignore the
details of the detector optics and simply treat the mirror as a {}``light
collector'' with a circular aperture. The focal plane pixels are
arranged on a rectangular grid with each pixel having a fixed angular
size. The pixel angular size is selected to meet the detector design
requirement of resolving a distance of 1~km on the surface. So, for
an orbit height $h$, the pixel angular size, $\delta$, is equal
to $1/h$~radian, with $h$ measured in km's. For a 800~km orbit
this translates to $\delta\approx0.7^{o}$. We assume full coverage
of the focal surface, i.e., we ignore the physical gaps and dead areas
between the PMT's.

During an event simulation we calculate and record the arrival time
of each photon reaching the detector from the shower. The arrival
direction of the photon determines the pixel in which the photon is
registered. Data for a triggered pixel comprises the pixel pointing
direction, the integrated pixel signal and the mean arrival time of
the photons recorded by the pixel. The integrated signal is simply
the number of photo-electrons (pe) recorded by the pixel in a time
window of 12~$\mu$s. The particular choice for the width of the
time window allows enough time for a shower with a zenith angle of
30$^{o}$ or greater to cross the field of view of the pixel.

We employ a simple detector trigger scheme, which requires at least
six pixels to fire from the light of the shower in order to form an
event. A pixel trigger occurs if the pixel records three or more pe
in a one $\mu$s interval. The test for an individual pixel trigger
is performed as follows:

\begin{enumerate}
\item the arrival times for each of the pe recorded by the pixel are sorted
in time to find the arrival times of the first and last recorded pe
A time window is formed around these times, and a 2~$\mu s$ interval
is added to each end.
\item sky noise pe are added to the expanded time window assuming a uniform
background of 200 photons/m$^{2}$/sr/ns. 
\item shower generated and noise pe are now stored in a histogram with a
1~$\mu$s bin width. The histogram's bins are scanned and if any
bin has three or more entries then the pixel trigger flag is set.
\item finally, if the total width of the time window exceeds 12~$\mu$s,
then a sliding window of that width is used to scan the histogram
to find the set of contiguous bins with the largest sum.
\end{enumerate}
Finally, we incorporate elements from the HiRes detector to cover
some of the gaps in the simulation. In particular we assume that the
detector will use a UV filter similar to that of HiRes, which passes
light in the 300-400~nm range. A parameterization of the wavelength
dependence of the PMT quantum efficiency is also borrowed from the
HiRes simulation, so is a constant mirror reflectivity of 80\%. These
detector components are described in detail in \cite{AbuZ_2k_nima}.

\subsection{\label{sub:The-Atmosphere-sec_Sim}The Atmosphere}

There are four elements or components to the simulation of UV light
transmission through the atmosphere. These include Rayleigh scattering
by molecules of air, scattering by surface aerosols, absorption by
ozone molecules, and scattering by clouds. The treatment of the first
three is based on the HiRes simulations; we adopted the same models
without modifications. Although these models are more appropriate
for the Utah desert observation conditions than for observation over
the ocean, the differences should have little effect on the results
of this study. This is because the shower development and light propagation
to the satellites occur almost entirely above the surface aerosol
layer, which is the one factor most likely to be significantly different
between the desert and the ocean. Before we turn to a discussion of
the cloud simulation we present a brief overview of these models.

Light scattering is characterized by the scattering cross section,
$\beta$, and the phase function\emph{,} $P=P(\cos\theta_{s})$, where
$\theta_{s}$ is the scattering angle. The cross section for molecular,
or Rayleigh, scattering is given by:

\begin{equation}
\beta_{R}=100(\rho(h)/x_{R})(400/\lambda)^{4}\label{eq:rayleigh_beta}\end{equation}
 where $\beta_{R}$ is measured in units of $m^{-1}$, $\rho(h)$
is the air density (g/cm$^{2}$) at altitude $h$ (m) above sea level,
and $x_{R}=2970$~g/cm$^{2}$ is the mean free path at wavelength
$\lambda=400$~nm. The air density and temperature profiles as function
of altitude are given by the U.S. Standard Atmosphere, 1976 \cite{Atmos76}.
The phase function for Rayleigh scattering is given by:

\begin{equation}
P(\cos\theta_{s})=(3/16\pi)\left(1+\cos^{2}\theta_{s}\right)\label{eq:rayleigh_phfunc}\end{equation}

\begin{figure}
\includegraphics[%
  scale=0.65]{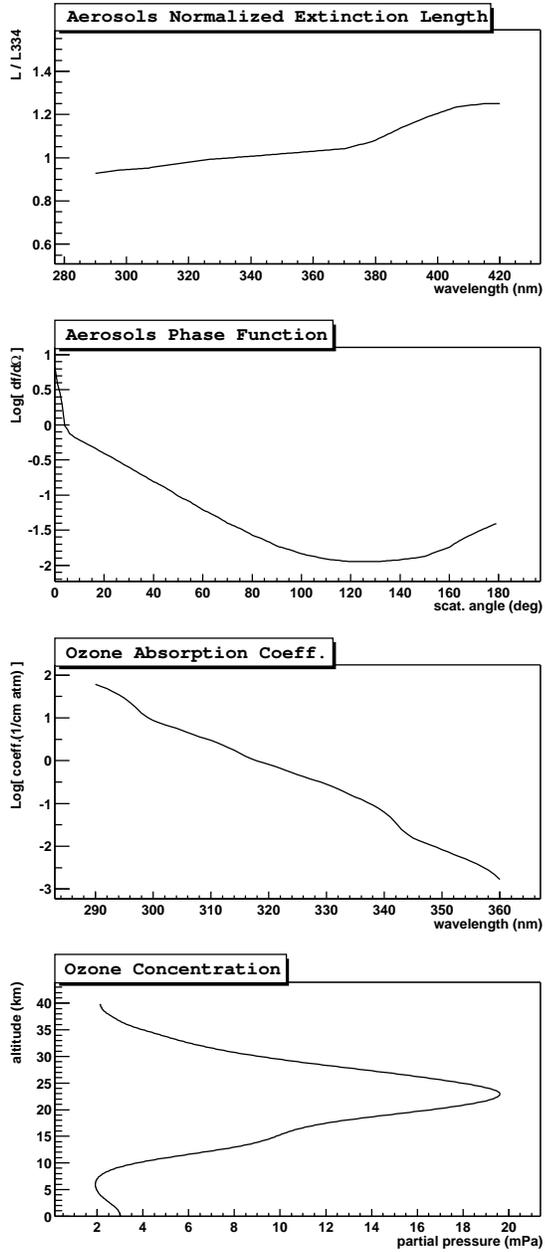}

\caption{\label{fig:mc97_atmos} From the top, the wavelength dependence of
the aerosols extinction length, the aerosols phase function, the wavelength
dependence of ozone absorption, and the ozone concentration as a function
of altitude.}
\end{figure}

Aerosols scattering is calculated according to the following formula:

\begin{equation}
\frac{dN}{dl}=-\frac{N}{L_{a}(\lambda)}\rho_{a}(h)\label{eq:aerosols_beta}\end{equation}
 Here $L_{a}$ is the scattering length at the surface, and we have:

\begin{equation}
\beta_{a}(\lambda)=\rho_{a}/L_{a}(\lambda)\label{eq:aerosols_beta_1}\end{equation}
The aerosols \emph{reduced} \emph{density,} $\rho_{a}$, is given
by:

\begin{equation}
\rho_{a}=\left\{ \begin{array}{cc}
1 & h<h_{m}\\
e^{-(h-h_{m})/H_{a}} & h\geq h_{m}\end{array}\right.\label{eq:aerosols_density}\end{equation}
 where $h_{m}$ is the height of the mixing layer, and $H_{a}$ is
the scale height above the mixing layer \cite{fe85}. The scattering
length at wavelength $\lambda=334$~nm is a free parameter of the
model. The wavelength dependence of the scattering process is accounted
for by a parameterization, shown in figure \ref{fig:mc97_atmos}.
The aerosols scattering phase function, also shown in figure \ref{fig:mc97_atmos},
is based on the Longtin desert aerosols model \cite{Longtin}. For
this study, the model parameters are set to: $h_{m}=0$, $H_{a}=1.2$~km,
and $L_{a}(334)=23.0$~km.

Ozone absorption is characterized by a wavelength-dependent absorption
coefficient, $\alpha_{O3}$, and an altitude-dependent concentration,
$\rho_{O3}$, both shown in figure \ref{fig:mc97_atmos}. Model parameters
were extracted from \cite{USAF_Handbook}. The extinction length,
in meters, due to ozone absorption can be written as: 

\begin{equation}
\frac{1}{L_{O3}}=(9.87\times10^{-7})\times\alpha_{O3}(\lambda)\times\rho_{O3}(h)\label{eq:ozone_ext}\end{equation}
 with the constant factor accounting for unit conversion.

The cloud model and the simulation of light propagation in a cloud
is described in detail in appendix \ref{sec:Clouds-Simulation}. Here
we present a brief description of the model.

Simulated clouds have a uniform density which steps to zero at the
cloud boundaries. The parameters used to describe a cloud are the
cloud base height, $h_{base},$ cloud top height, $h_{top}$, and
the optical depth $\tau$. The scattering length, $\beta$, inside
the cloud is related to the optical depth by the relation:

\begin{equation}
\tau=\beta(h_{top}-h_{base})\label{eq:cloud_tau}\end{equation}
 The wavelength dependence of the scattering process, which is mild
in the near UV \cite{Liou_fig5.3}, is ignored. Finally, scattering
within optically thin clouds calculatioins use a phase function appropriate
for ice clouds at $\lambda=0.5$~$\mu$m \cite{Liou_p276}.

\subsection{\label{sub:Shower-Simulation}Shower Simulation}

The primary particle energy is selected at the start of the simulation.
The shower track geometry is generated randomly in order to obtain
uniform and isotropic showers distributions in the atmospheric volume
viewed by the detector.

The generated shower profiles follow a Gaisser-Hillas function \cite{GaisserHillas}.
The number of electrons, $N_{e}$, in the shower is given as a function
of depth, $x$, by:

\begin{equation}
N_{e}(x)=N_{max}\left(\frac{x-x_{0}}{x_{max}-x_{0}}\right)^{(x_{max}-x_{0})/\lambda}e^{(x_{max}-x)/\lambda}\label{eq:GH}\end{equation}
where the $x_{0}$ parameter is chosen from an exponential ($<x_{0}>=$35~g/cm$^{2}$
for proton primaries), $x_{max}$ is chosen from a Gaussian distribution,
$\lambda$ is fixed at 70 gm/cm$^{2}$, and $N_{max}$ is selected
so that the integral of the profile, corrected for lost energy \cite{Linsley83}
\cite{Song_2k}, gives a total shower energy equal to that of the
primary particle:

\begin{equation}
E_{tot}=E_{corr}+2.18\times\int N_{e}dx\label{eq:Etot}\end{equation}

The mean and variance of the $x_{max}$ Gaussian distribution depend
on the energy and mass number of the primary cosmic ray particle.
For protons we assume $<x_{max}>=725$~g/cm$^{2}$ at $E=10^{18}$~eV,
increasing by 55~g/cm$^{2}$ per decade in energy. The standard deviation
is set to 50~g/cm$^{2}$ for all energies. These values are based
on shower simulations quoted by the Fly's Eye group in their analysis
\cite{BirdPRL93}, and are consistent with simulations results from
the \noun{corsika} program with the \noun{}QGSJet model \cite{Pryke_2k1}.

The Nishimura-Kamata-Greisen (NKG) function \cite{NKG_NK,NKG_G},
is used to describe the lateral distribution of shower electrons:

\[
\rho_{e}(r)=\frac{N}{r^{2}}f\left(s,\frac{r}{r_{M}}\right)\]
where

\begin{equation}
f=\frac{\Gamma(4.5-s)}{2\pi\Gamma(s)\Gamma(4.5-2s)}\left(\frac{r}{r_{M}}\right)^{s-2}\left(1+\frac{r}{r_{M}}\right)^{s-4.5}\label{eq:NKG}\end{equation}
 and the shower age parameter, $s$, is given by:

\begin{equation}
s=3x/(x+2x_{max})\label{eq:shower_age}\end{equation}
Here, $r_{M}$ is the Moli\`{e}re radius. The value of $r_{M}$ depends
on the air density and is evaluated at each point along the track.

Fluorescence light is generated according to the formulas given in
\cite{fe85}, but more recent measurements of the air fluorescence
yield are used \cite{Loh}. The calculation of the \v{C}erenkov light
production also follows that of the Fly's Eye paper \cite{fe85}. 

The above procedure is modified if optically thin clouds are present.
First we identify all track segments which lie \emph{inside} the cloud.
For these segments the simulation proceeds as described above but
in addition, the number of fluorescence photons scattering once in
the cloud, and \v{C}erenkov photons scattering once or twice in the
cloud, before reaching the detector, are calculated. These photons
are included in the detector response as additional signal.

\subsection{\label{sub:Cerenkov-Spot-sec_Sim}\v{C}erenkov Spot}

The \v{C}erenkov spot refers to the area on the surface around the
shower core where the \v{C}erenkov beam is reflected, off the surface
and into the detector. The simulation of the signal recorded by the
detector, and generated by the reflected beam, requires the consideration
of three separate issues: (1) The lateral distribution of the \v{C}erenkov
photons at a given point along the shower development, (2) the lateral
distribution of the \v{C}erenkov photons at the surface, and (3)
the reflection properties of the surface.

The lateral distribution of photons in the \v{C}erenkov beam is assumed
to follow that of the shower electrons, i.e. it's given by the NKG
function. This is a simplification and in general results in a greater
concentration of \v{C}erenkov photons near the shower axis. A more
accurate description of the lateral spread of the \v{C}erenkov beam
would be required ( along with a detailed detector simulation ) to
address the problem of identifying cloud presence from event data.
For this study, however, the use of the NKG function is sufficient.

The \v{C}erenkov front has a circular shape centered around, and
perpendicular to the shower axis. The spot formed on the surface by
the beam is in general elliptical with the elongation of the spot
depending on the zenith angle of the shower. In the simulation, the
transformation of the photon position from a point on a circular disk
about the shower axis to a point on the reflecting surface is performed
and the time offset of the photons relative to the shower core is
calculated before the photon is propagated to the detector.

Finally, the surface reflection albedo is assumed to be constant at
20\% for reflection off water, and the reflection is assumed to be
isotropic (into $2\pi$). The same is assumed for cloud reflection.

\section{\label{sec:Event-reconstruction}Event Reconstruction}

Events are reconstructed from the raw event data to obtain the shower
energy, shower $x_{max}$ , and the arrival direction of the primary
cosmic ray particle. The raw data consists of a set of triggered pixels
with known pointing directions, each with a measured mean arrival
time and a time-integrated total pe count. With two instruments observing
the shower, the combined data from each instrument comprises a \emph{stereo
event.} Data from each instrument can be analyzed separately as a
\emph{monocular} \emph{event}.

The reconstructed shower is described by a set of parameters which
specify the shower geometry and shower profile. The shower energy
is obtained from the shower profile using eq. \ref{eq:Etot}. The
shower track geometry can be described by a pair of orthogonal vectors
$\vec{R}_{p}$ and $\hat{u}_{t}$, the latter being the shower direction
unit vector. Alternatively the geometry can be specified in terms
of the Shower-Detector (SD) plane normal, $\hat{n}$, and a pair of
scalars $R_{p}$ and $\psi$ which determine a line in that plane.
The shower profile is given by the Gaisser-Hillas function, eq. \ref{eq:GH}. 

Event reconstruction is divided into three consecutive steps:

\begin{enumerate}
\item SD plane reconstruction for each eye.
\item Shower track geometry reconstruction.
\item Shower profile and energy reconstruction.
\end{enumerate}
Step 2. above is implemented differently for monocular and stereo
events. For stereo, the intersection of the two SD planes from each
eye determines the shower track. Monocular reconstruction requires
the use of pixel trigger timing and an additional constraint provided
by the observation of the \v{C}erenkov spot. All other steps are
similar for both monocular and stereo events.

In general, a file containing a set of Monte Carlo generated events
contains a reference to the set of atmospheric parameters used in
the simulation. This enables the reconstruction programs to use the
same atmosphere used in the simulation. However, since the purpose
of this study is to investigate the effect of clouds which go undetected,
on the event reconstruction, clouds are \emph{removed} from the atmosphere
during reconstruction.

\subsection{\label{sub:Shower-Geometry}Shower Geometry}

\subsubsection{Shower-Detector Plane}

The shower-detector plane is that plane which contains the detector,
\emph{a point}, and the shower track, \emph{a line}. For a stereo
detector a SD plane is calculated separately for each eye. The SD
plane is calculated by minimizing a $\chi^{2}$ function given by:

\begin{equation}
\chi^{2}=\sum_{i}\frac{[(\hat{n}\cdot\hat{n}_{i})]^{2}\cdot w_{i}}{\sigma_{i}^{2}}\end{equation}
 where the sum is over triggered pixels, $\hat{n}$ is the plane normal,
$\hat{n}_{i}$ are the pixel viewing direction vectors and $w_{i}$
are weights equal to the total number of pe seen by pixel $i$. An
angular pointing error of $\sigma\sim0.07^{o}$ (equal to the pixel
angular size) is assumed for all pixels.

\subsubsection{Shower Track in the SD plane}

In the case of stereo observation, the intersection of the SD plane
normals from each eye describes a line in space, namely the shower
track. This method despite its simplicity works very well in general
\cite{Wilkinson_99}. Only events for which the opening angle between
the two planes is small and the plane determination was not good,
e.g. due to short track-length, does the method fail to produce accurate
results.

In the case of monocular observation, track reconstruction uses the
pixel timing information. The timing fit method is based on the relation
between the crossing time of the shower front in a pixel's field of
view (mean photon arrival time at the detector) and the pixel's viewing
angle. The pixel crossing time $t_{i}$ as a function of the pixel
viewing angle $\chi_{i}$ in the SD plane, is given by:

\begin{equation}
t_{i}=t_{0}+\frac{R_{p}}{c}tan\frac{1}{2}(\pi-\psi-\chi_{i})\label{eq:time_fit}\end{equation}
The reader can refer to \cite{fe85} for a derivation and definition
of the parameters. In all, there are three unknown fit parameters,
namely $t_{o}$ , $R_{p}$, and $\psi$.

A fit based on eq. \ref{eq:time_fit} produces accurate results only
when the range of angles covered by the triggered pixels, i.e. the
angular track-length of the event, is {}``large enough''. A good
discussion of the timing relation and the requirements for accurate
reconstruction appear in \cite{Sommers_95} in relation to the Auger
detector. For showers observed from space, the angular track-length
is too small for the fit to result in satisfactory results, and an
additional constraint on the shower geometry is required. The observation
of the \v{C}erenkov spot provides this constraint in the form of
a known shower impact point on the surface, also referred to as the
\emph{shower} \emph{core} position. The use of this constraint results
in a significant improvement in the accuracy of the fit. The shower
core vector is obtained from the event data as follows:

If the reflected \v{C}erenkov light is observed by one pixel with
a pointing direction vector $\hat{v}_{c}$ then, the shower core vector
can be calculated using the relation $\vec{r}_{c}=\vec{r}_{m}+s\hat{v}_{c}$
where $\vec{r}_{c}$ is the shower core position, $\vec{r}_{m}$ is
the detector position vector, and $s$ is a scalar which can be solved
for by making the requirement that this line intersect the surface
of the Earth (a sphere with a known radius corresponding to an altitude
of $h=0$~m above sea level). 

In general, the reflected light is observed by one or more pixels,
to identify which we examine the set of triggered pixels for the one
that triggered last in time and the one triggered last in angular
distance from the start of the track. A group of one or more pixels
is first identified as being triggered by the reflected beam by examining
those pixels adjacent to the last triggered pixel for their trigger
times. A sum of the pointing directions of these pixels (weighted
by the total signal in each pixel) is performed to get an average
direction. This direction is then projected in the SD plane to get
a final estimate of the core direction.

\subsection{\label{sub:profile-reconstruction}Shower Profile}

With the shower track geometry in hand we proceed to reconstruct the
shower profile. The shower profile is assumed to follow a Gaisser-Hillas
function with three free parameters: $x_{0}$, $x_{max}$, and $N_{max}$.
For each trial profile a shower is generated with the reconstructed
geometry and the detector response to the shower is calculated. The
calculation proceeds along the same lines as the Monte Carlo (same
light production and propagation models) with the exception that all
random fluctuations are suppressed. Where in the MC the number and
starting position of ray-traced photons is chosen randomly; During
reconstruction, the mean number of photons from each track segment
is distributed on a two dimensional grid representing an NKG lateral
distribution. The effect of the finite mirror spot size is accounted
for by distributing the flux received by the detector among the pixels
according to the distribution of the spot.

The best fit (reconstructed shower profile parameters) is chosen to
minimize a $\chi^{2}$ function calculated from the observed (MC output)
and fit values for the pe counts from each pixel in the event. The
following function is used:

\begin{equation}
\chi_{pfl}^{2}=\sum_{i}\frac{1}{\sigma_{i}^{2}}(S_{i}^{(m)}-S_{i}^{(p)})^{2}\label{eq:chi2_pfl}\end{equation}
 where the sum is over triggered pixels, $S_{i}^{(m)}$ is the measured
pixel signal in pe, $S_{i}^{(p)}$ is the predicted pixel signal,
and $\sigma_{i}^{2}=S_{i}^{(m)}+B_{i}$. The $\sigma_{i}^{2}$ terms
are obtained by adding in quadrature the Poisson fluctuation in the
signal, $\sqrt{S^{(m)}}$, and the estimated sky background fluctuations
for that pixel. It should be noted that not all triggered pixels are
included in the sum. Those pixels near the end of the track believed
to be triggered by the reflected \v{C}erenkov beam are excluded from
the sum.

\section{\label{sec:Results}Results and Conclusions}

In this section we summarize the results obtained from the two studies
of the possible effects of cloud presence on the reconstructed shower
energies and $x_{max}$. We start with optically thin clouds.

\subsection{\label{sub:Thin-Clouds-sec_Results}Optically Thin Clouds}

The effect of cloud presence is examined as follows: A Monte Carlo
shower is generated in a cloud free atmosphere and the detector response
is evaluated. Next, a loop over a set of cloud configurations (described
below) is made in which the selected cloud is included in the atmosphere
simulation. The same shower from the cloud free simulation is developed
through the atmosphere, and the detector response is recorded. All
events are reconstructed assuming a cloud free atmosphere.

Of the many (infinite) possible cloud configurations we selected the
following set: Cloud base height is set to 6~km for all clouds in
the study. Four different cloud top heights are used, these are given
in table \ref{table:slant_vs_h} along with the corresponding atmospheric
slant depths along shower tracks at 60$^{o}$ and 75$^{o}$. At each
cloud top height setting, the cloud optical depth is varied between
0.1-0.5 in steps of 0.1, for a total of 20 cloud settings. Most Proton
initiated showers with an energy of $10^{21}$~eV are expected to
have $x_{max}$values in the range of 800-1000~g/cm$^{2}$. If these
showers develop at a zenith angle of $60^{o}$ then they will reach
maximum development at an altitude just above the selected cloud base
height. In most cases then the shower will traverse the cloud while
it is still increasing in size. At $\theta=75^{o}$, A cloud top height
of 7.28~km or 9.25~km insures that the cloud lies below the shower
$x_{max}$. The other cases cover a larger portion of the shower development
curve.

\begin{table}

\caption{\label{table:slant_vs_h} slant depth (gm/cm$^{2}$) vs. height (km)
for showers with zenith angles of $60^{o}$and 75$^{o}$. The values
shown in the first column are chosen}

\vspace{1ex}

\begin{tabular}{|c|r|r|}
\hline 
$h$ (km) \textbackslash{} $\theta$&
60$^{o}$&
75$^{o}$\tabularnewline
\hline
\hline 
6.00&
956&
1810\tabularnewline
\hline
\hline 
7.28&
800&
1512\tabularnewline
\hline 
9.25&
600&
1130\tabularnewline
\hline 
11.85&
400&
751\tabularnewline
\hline 
16.23&
200&
372\tabularnewline
\hline
\end{tabular}
\end{table}

\begin{figure}
\includegraphics[%
  scale=0.65]{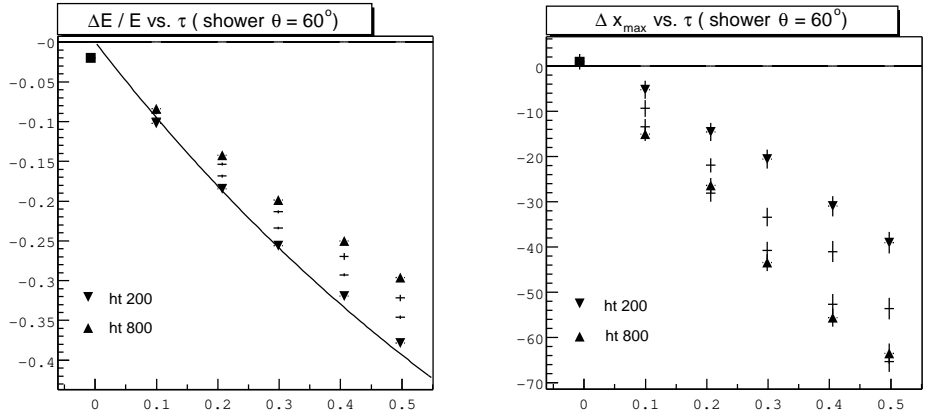}

\caption{\label{fig:Results_thin_stereo_60}Energy and $x_{max}$ shift in
the presence of optically thin clouds ( $\tau$ plotted along the
$x$-axis ) for four different cloud top height settings, indicated
by {}``ht~200'' through {}``ht~800'' . Stereo reconstruction
of showers with $\theta=60^{o}$. The function, $\exp(-\tau)-1$,
is superimposed on the energy plot.}
\end{figure}

\begin{figure}
\includegraphics[%
  scale=0.65]{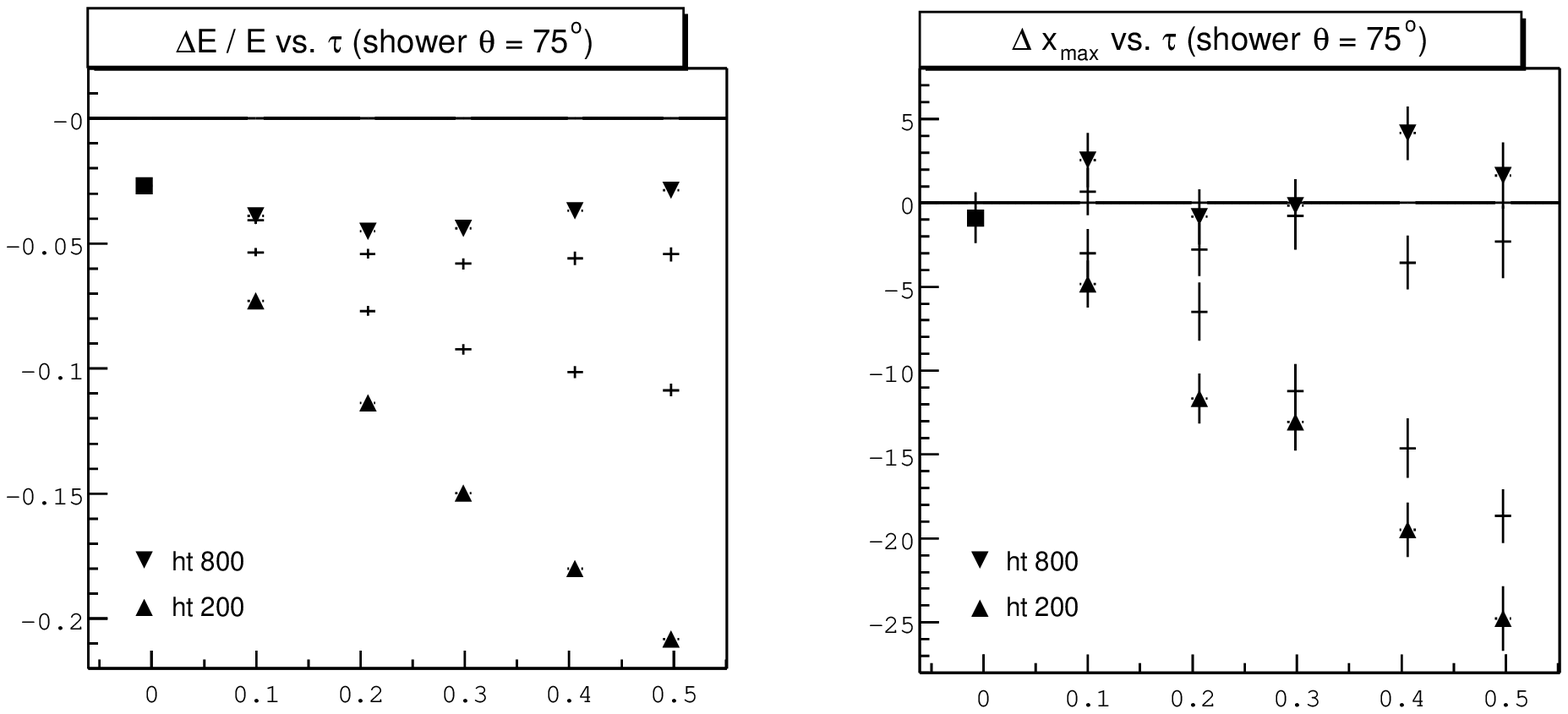}

\caption{\label{fig:Results_thin_stereo_75} Energy and $x_{max}$ shift in
the presence of optically thin clouds ( $\tau$ plotted along the
$x$-axis ) for four different cloud top height settings, indicated
by {}``ht~200\char`\"{} through {}``ht~800\char`\"{}. Stereo reconstruction
of showers with $\theta=75^{o}$.}
\end{figure}

\begin{figure}
\includegraphics[%
  scale=0.65]{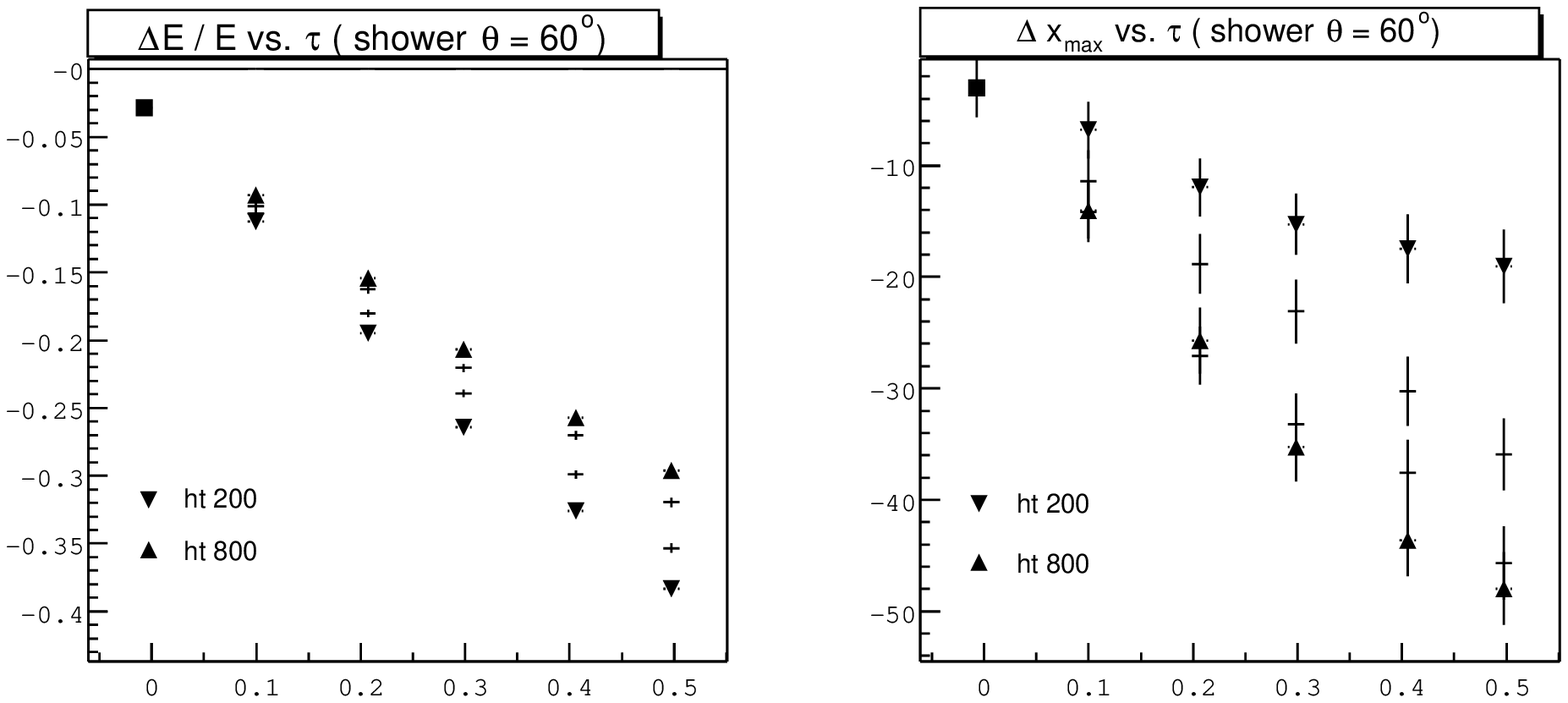}

\caption{\label{fig:Results_thin_mono_60} Energy and $x_{max}$ shift in
the presence of optically thin clouds ( $\tau$ plotted along the
$x$-axis ) for four different cloud top height settings, indicated
by {}``ht~200\char`\"{} through {}``ht~800\char`\"{}. Monocular
reconstruction of showers with $\theta=60^{o}$.}
\end{figure}

\begin{figure}
\includegraphics[%
  scale=0.65]{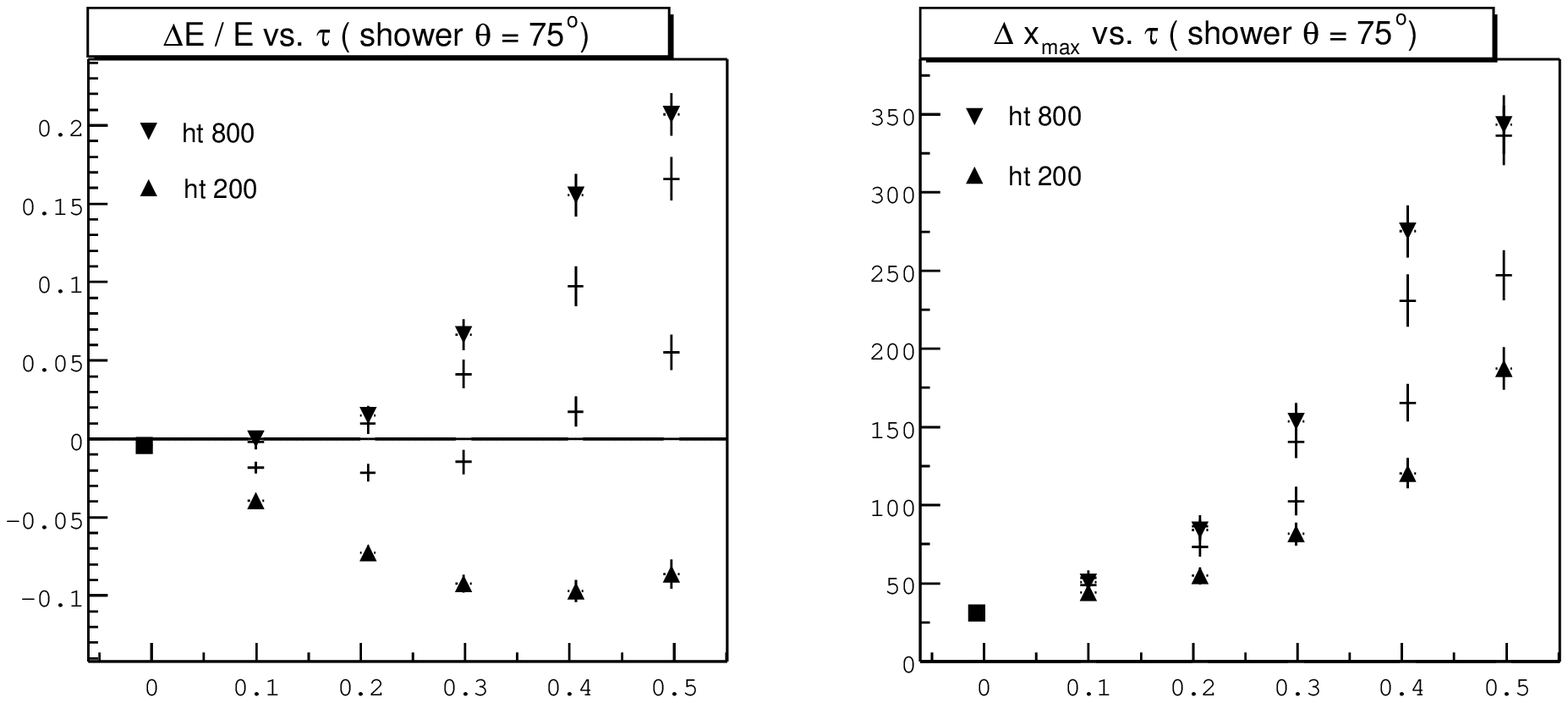}

\caption{\label{fig:Results_thin_mono_75} Energy and $x_{max}$ shift in
the presence of optically thin clouds ( $\tau$ plotted along the
$x$-axis ) for four different cloud top height settings, indicated
by {}``ht~200\char`\"{} through {}``ht~800\char`\"{}. Monocular
reconstruction of showers with $\theta=75^{o}$.}
\end{figure}

Figures \ref{fig:Results_thin_stereo_60} through \ref{fig:Results_thin_mono_75}
summarize the results for stereo and monocular reconstruction of shower
energy and $x_{max}$. In the figures, the cloud optical depth, $\tau$,
is plotted along the $x$-axis, with $\tau=0$ corresponding to the
cloud free atmosphere. Four points for each $\tau>0$ correspond to
the different $h_{top}$ values. Cloud $h_{top}$ values of 7.28~km
and 16.23~km are explicitly indicated on the plots by their corresponding
slant depth values at $\theta=60^{o}$. Each point in the plots represents
the mean shift and standard deviation for a set of 200 reconstructed
events. 

In the case of stereo reconstruction, the reconstructed shower geometry
is unaffected by cloud presence. The cloud affects the amount of light
reaching the detector from different parts of the shower depending
on its position and extent. In a couple of cases the effect can be
easily understood. For example, the case of $h_{top}=16.23$~km (ht~200
in the plot) and shower $\theta=60^{o}$, the mean reconstructed energy
is shifted down by a factor of $\exp(-\tau)$. Another example is
provided by the $\theta=75^{o}$ showers when the cloud lies well
below shower maximum ($h_{top}$=~7.28~km or 9.25~km), in these
cases we see very little effect on the reconstructed shower parameters. 

The dependence of monocular shower geometry reconstruction on the
observation and correct identification of the \v{C}erenkov spot makes
the interpretation of the results more complicated. In general, if
the shower geometry is reconstructed correctly, then the effect of
the cloud on the reconstructed energy and $x_{max}$ will be similar
to the effect on stereo reconstructed events. Otherwise, the error
will depend mostly on what the reconstructed geometry turns out to
be. Looking at the $\theta=60^{o}$ showers, one can see that the
energy plot looks almost identical to that of the stereo case, and
that the mean shift in $x_{max}$ is comparable but smaller than that
for the stereo case, however, the error bars are slightly larger in
the monocular case. For most showers in this group, the detector did
trigger on the reflected Cerenkov light and the geometry reconstruction
procedure gave the right results. In a few cases, as the cloud optical
depth increased, the detector did not trigger on the reflected beam
and a wrong geometry resulted based on the false identification of
the ground spot.

\begin{figure}
\includegraphics[%
  scale=0.6]{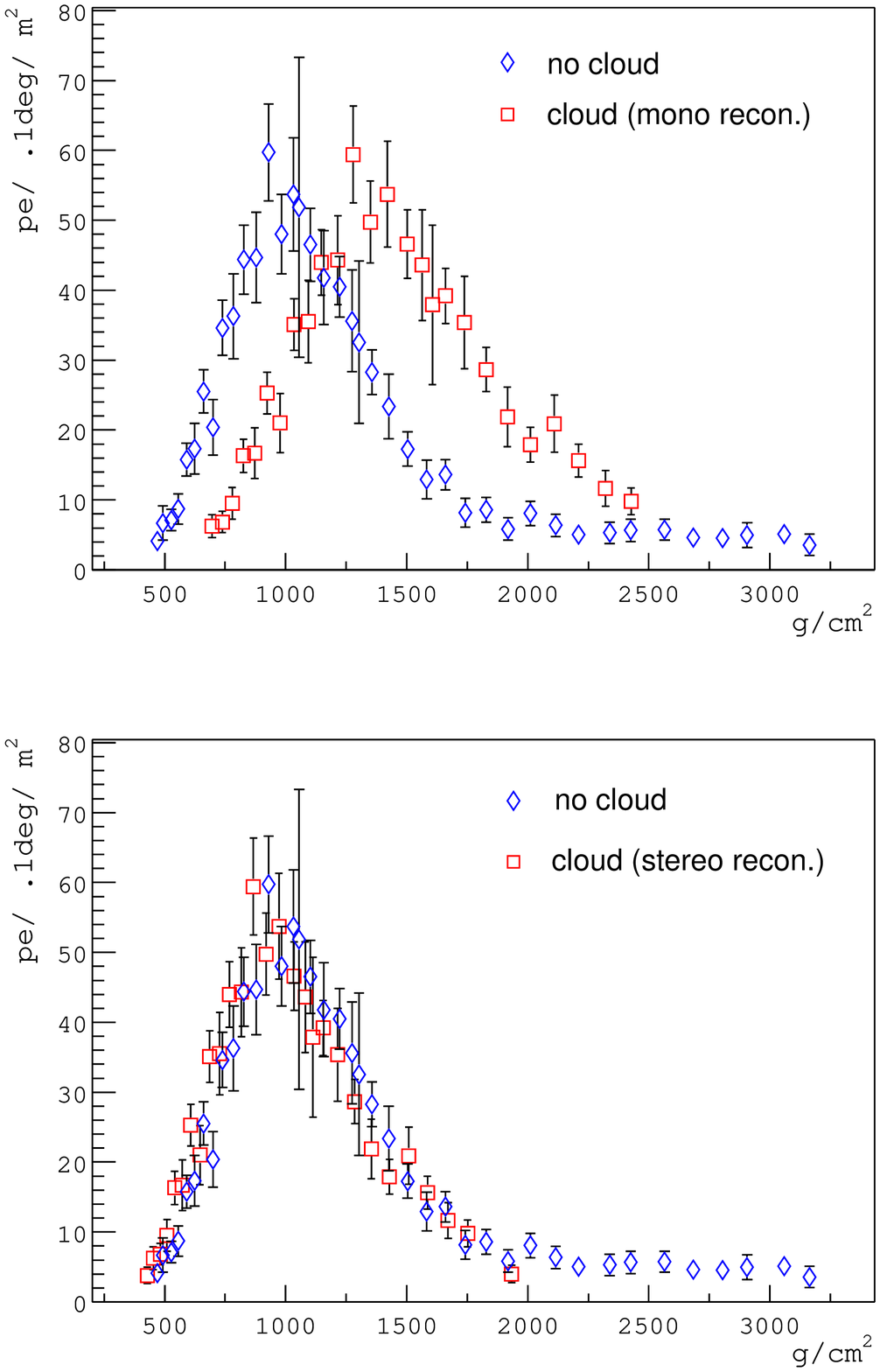}

\caption{\label{fig:Results-thin-Example1}Example showing the effect of a
cloud on the calculated light flux (pe/0.1deg/m$^{2}$) seen by the
detector. In this example a 10$^{21}$~eV proton shower with a zenith
angle of $75^{o}$ passes through a cloud with an optical depth of
0.4 which extends between 6 km ($\approx1500$~g/cm$^{2}$) and 7.28~km
($\approx1800$~g/cm$^{2}$) }
\end{figure}

For showers with $\theta=75^{o}$, the true Cerenkov spot was not
observed in a large number of cases. This can be explained by noting
that: (a) the larger inclination of the shower means that the Cerenkov
beam will go through a larger distance through the cloud resulting
in more attenuation, and (b) especially for the cases in which the
cloud lies below the shower maximum development, there are no more
shower particles to feed the \v{C}erenkov beam. Figure \ref{fig:Results-thin-Example1}
shows an example of a reconstructed event in the presence of a cloud
with $\tau=0.4$ lying below the shower development. 

\subsection{\label{sub:Thick-Clouds-sec_Results}Optically Thick Clouds}

For this study we place optically thick clouds with $h_{top}$ in
the range of 1-5~km, in one km steps, and generate showers at a fixed
energy of 10$^{21}$~eV and with fixed zenith angles of 60$^{o}$
and 75$^{o}$. Sets of 500 events each were generated for each combination
of shower zenith angle, and cloud height. All the data sets were reconstructed
and analyzed assuming no clouds. A number of quality cuts were applied
to the reconstructed sets of showers in order to remove badly reconstructed
events. A list of the applied cuts follows:

\begin{enumerate}
\item $\chi_{pfl}^{2}/ndof<10$ 
\item SD planes opening angle > 6$^{o}$. (Stereo events)
\item Observed angular track-length > 0.6$^{o}$ (approx. 9 pixels)
\item Number of good angular bins > 5 
\item $x_{max}$ bracket cut: $x_{first}+100<x_{max}$.
\end{enumerate}
Cuts number 2 and 3 remove events where the geometrical reconstruction
results were probably not accurate. Cut number 4 is somewhat similar
to 3, but is over \emph{good} angular bins. The last cut represents
the requirement that the shower $x_{max}$ was observed. Only events
which passed the cuts are included in the results.

\begin{figure}
\includegraphics[%
  scale=0.35]{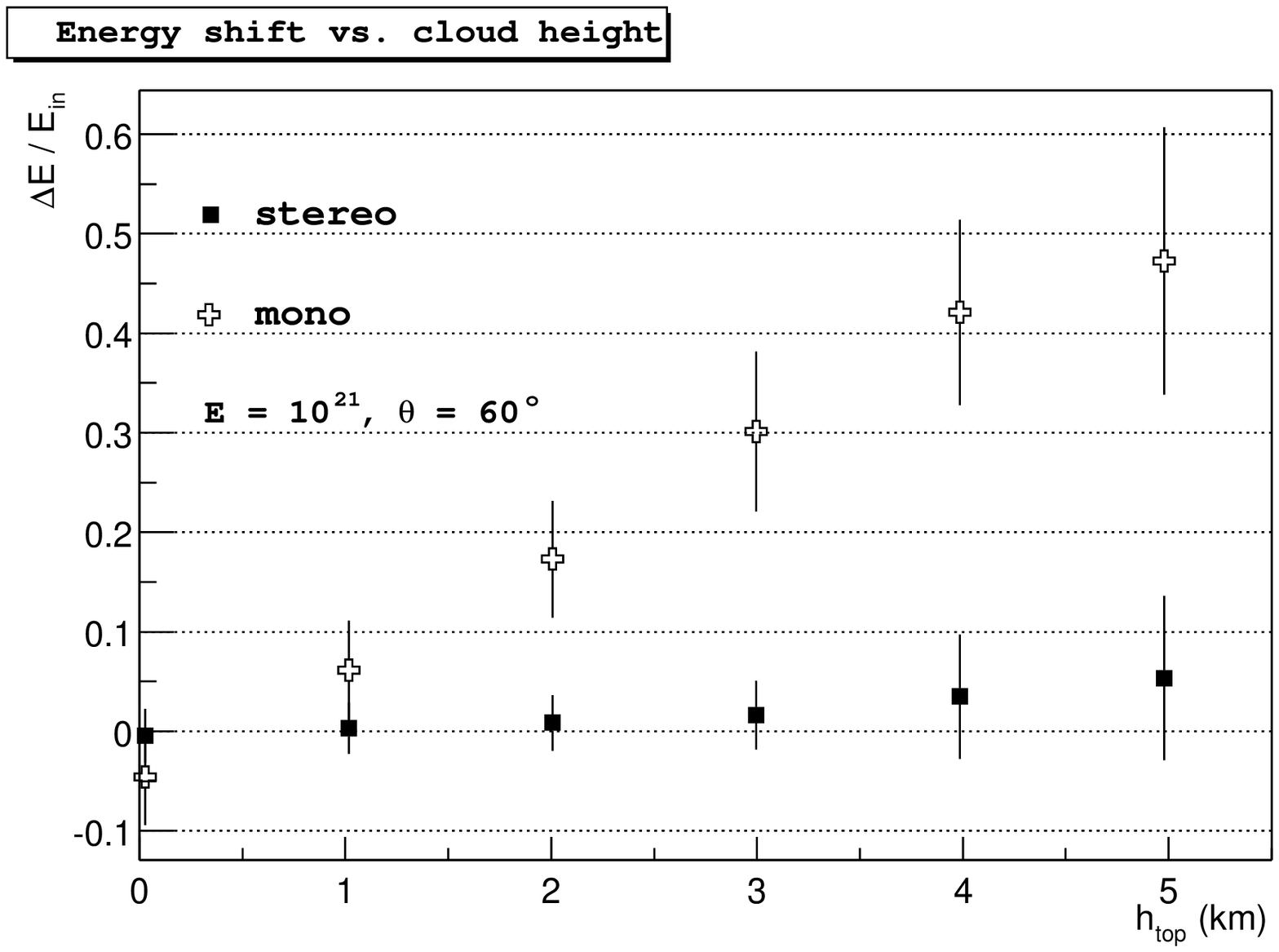}\includegraphics[%
  scale=0.35]{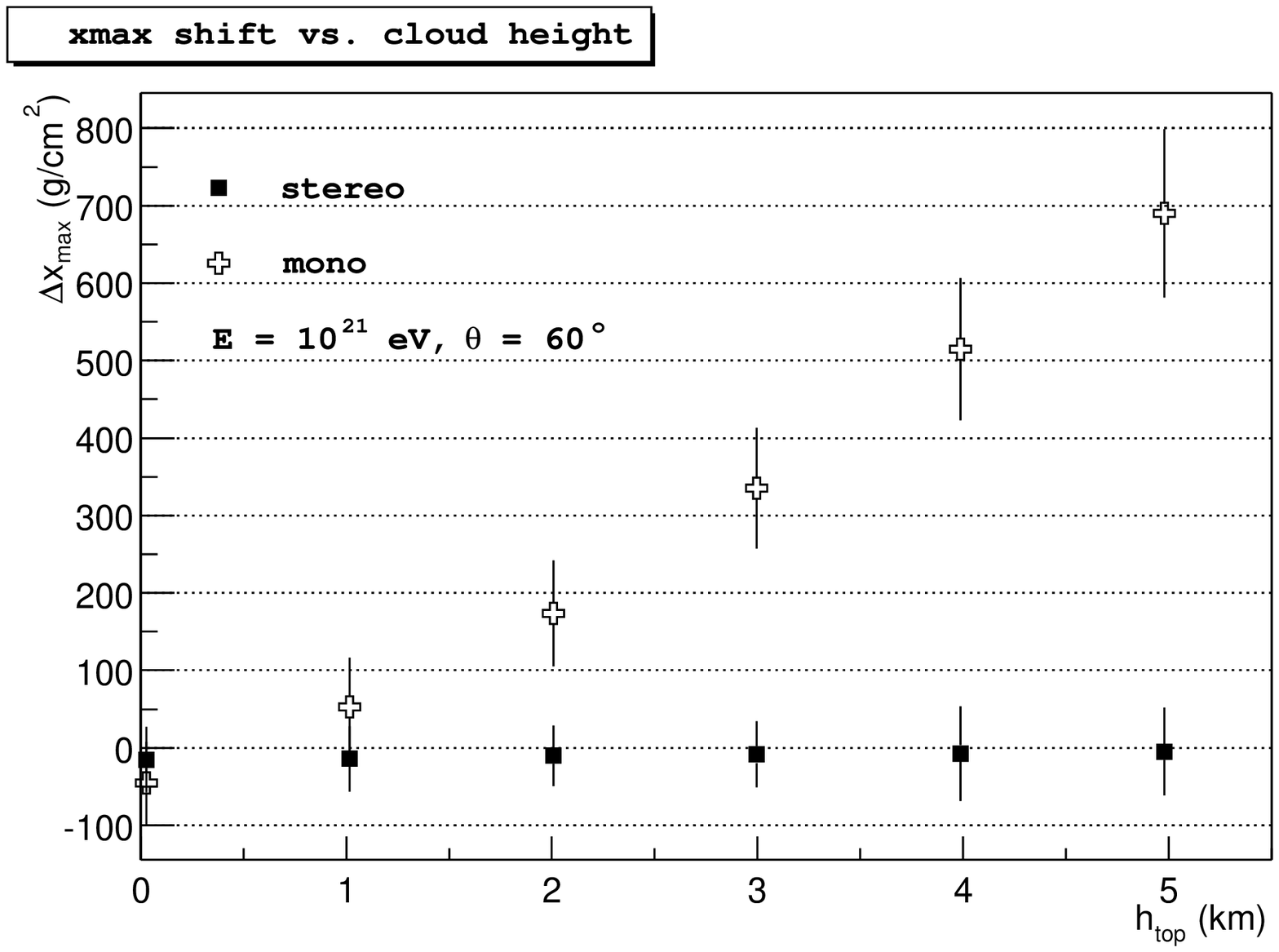}

\caption{\label{fig:Results-th60-Energy-Shift}Energy and $x_{max}$ shift
for 10$^{21}$~eV showers with zenith angle $\theta=60^{o}$ as a
function of cloud top height.}
\end{figure}

\begin{figure}
\includegraphics[%
  scale=0.35]{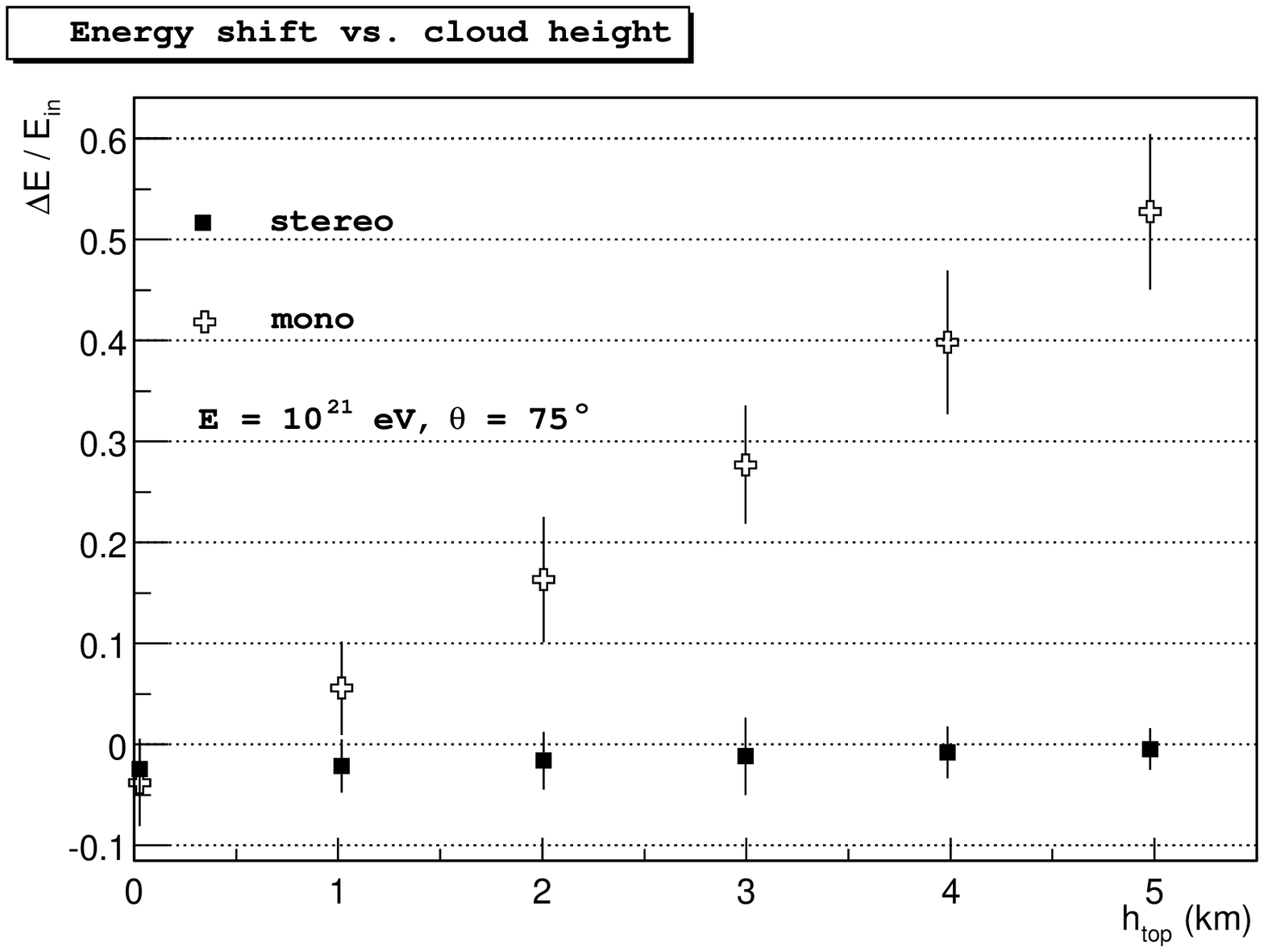}\includegraphics[%
  scale=0.35]{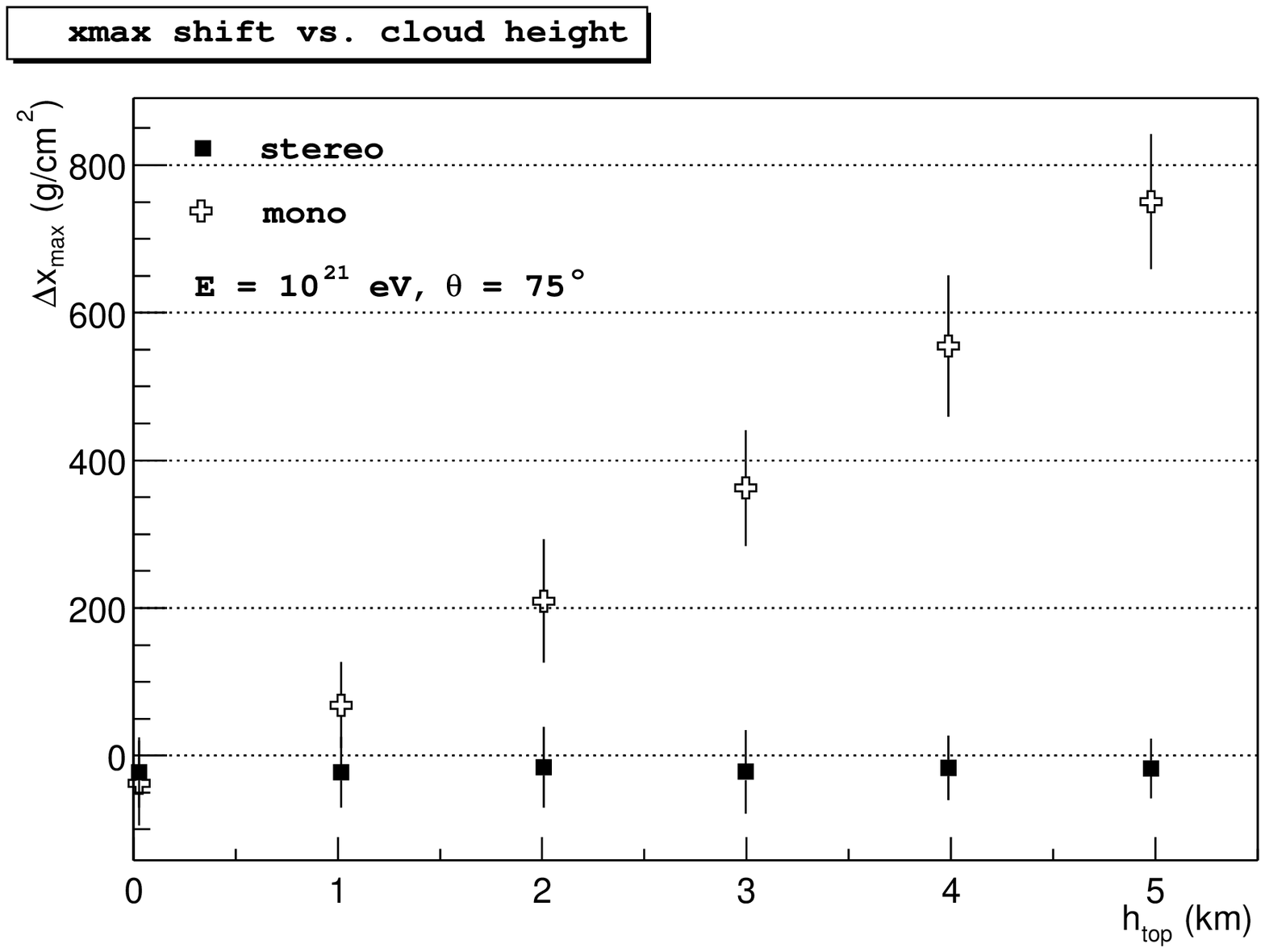}

\caption{\label{fig:Results-th75-Energy-Shift}Energy and $x_{max}$ shift
for showers with $\theta=75^{o}$and $E=10^{21}$eV. The x-axis is
the cloud top height in km's and the y-axis is $(E_{out}-E_{in})/E_{in}$. }
\end{figure}

Figure \ref{fig:Results-th60-Energy-Shift} shows the resulting shift
in energy and $x_{max}$, for the case $\theta=60^{o}$. Figure \ref{fig:Results-th75-Energy-Shift}
shows the same for $\theta=75^{o}$. The results are also shown in
tables \ref{tab:Results-th60-Energy-shift}, and \ref{tab:Results-th75-Energy-Shift}.
The \#events in these tables indicates the number of successfully
reconstructed events, out of 500. For the set $\theta=75^{o}$ and
clouds at 4~km the reconstruction job terminated before the full
set was done, and therefore, the smaller number of events.

\begin{table}

\caption{\label{tab:Results-th60-Energy-shift}Energy shift (\%) versus cloud
height for shower energy of $10^{21}$~eV and shower zenith angle
of 60$^{o}$. First column is cloud height in km. the following two
are for stereo while the last two are for monocular.\vspace{1ex}}

\begin{tabular}{|c|c|c|c||c|c|c|}
\hline 
$h_{top}$ &
stereo&
 $E$ shift&
$x_{max}$ shift&
mono&
$E$ shift&
$x_{max}$ shift\tabularnewline
(km)&
\#events&
\%&
g/cm$^{2}$&
\#events&
\%&
g/cm$^{2}$\tabularnewline
\hline
\hline 
0&
410&
-0.4&
-0.0&
471&
-4.6&
-41.7\tabularnewline
\hline 
1&
422&
0.3&
1.4&
472&
6.0&
75.7\tabularnewline
\hline 
2&
384&
0.9&
3.7&
458&
17.2&
200.4\tabularnewline
\hline 
3&
385&
1.9&
5.1&
458&
30.8&
351.0\tabularnewline
\hline 
4&
388&
3.0&
6.6&
446&
41.4&
500.2\tabularnewline
\hline 
5&
386&
3.8&
5.6&
374&
47.1&
644.0\tabularnewline
\hline
\end{tabular}
\end{table}

\begin{table}

\caption{\label{tab:Results-th75-Energy-Shift}Energy shift (\%) versus cloud
height for shower energy of $10^{21}$~eV and shower zenith angle
of 75$^{o}$. First column is cloud height in km. the following two
are for stereo while the last two are for monocular.\vspace{1ex}}

\begin{tabular}{|c|c|c|c||c|c|c|}
\hline 
$h_{top}$ &
stereo&
 $E$ shift&
$x_{max}$ shift&
mono&
$E$ shift&
$x_{max}$ shift\tabularnewline
(km)&
\#events&
\%&
g/cm$^{2}$&
\#events&
\%&
g/cm$^{2}$\tabularnewline
\hline
\hline 
0&
411&
-2.2&
-3.3&
469&
-3.9&
-21.7\tabularnewline
\hline 
1&
411&
-1.9&
-3.5&
470&
5.9&
101.7\tabularnewline
\hline 
2&
419&
-1.5&
-1.3&
461&
16.5&
240.3\tabularnewline
\hline 
3&
391&
-1.3&
-3.6&
454&
28.0&
389.1\tabularnewline
\hline 
4&
229&
-0.9&
-1.0&
254&
40.7&
553.5\tabularnewline
\hline 
5&
400&
-0.4&
-1.2&
293&
53.1&
727.5\tabularnewline
\hline
\end{tabular}
\end{table}

\begin{figure}
\includegraphics[%
  scale=0.4]{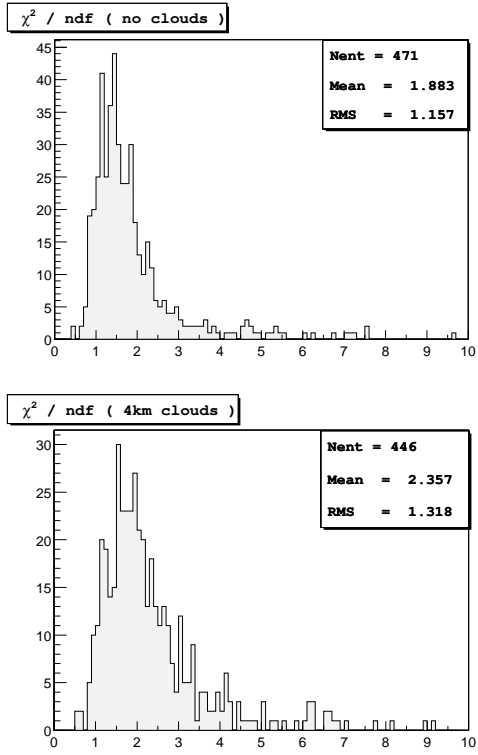}

\caption{\label{fig:Results-chi2-mono}$\chi^{2}/ndf$ distribution from monocular
reconstruction of two sets of showers: one set with no clouds, the
other with clouds at 4~km. Showers were generated with a fixed energy
of 10$^{21}$~eV and a zenith angle of 60$^{o}$.}
\end{figure}

The results show that while the performance of stereo is stable for
different cloud heights, monocular reconstruction suffers badly if
clouds at altitudes of 2~km or higher are present and their presence
goes unrecognized. The reconstructed shower profiles in the monocular
case {}``look'' normal with the exception of the abnormal development
depth and result in reasonable values for the $\chi^{2}$ as shown,
for an example, in figure \ref{fig:Results-chi2-mono}. 

\begin{figure}
\includegraphics[%
  scale=0.5]{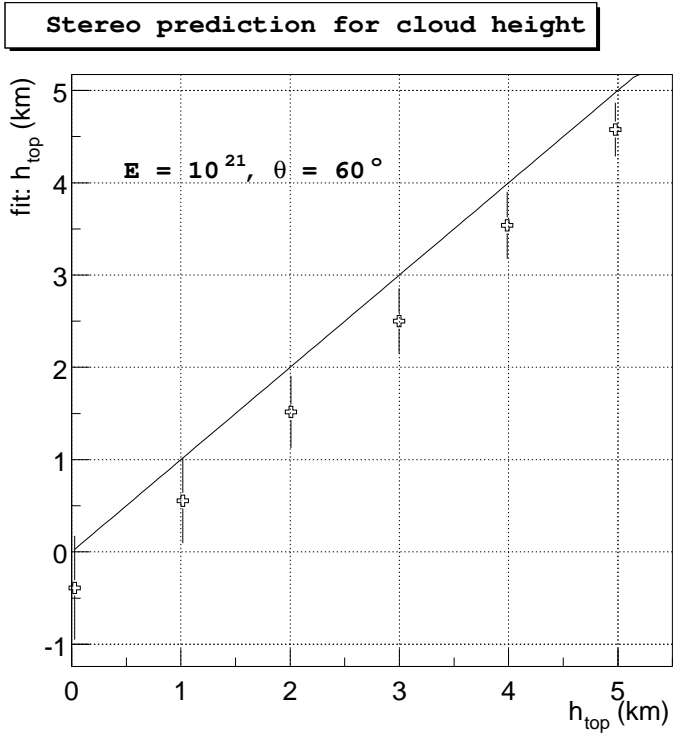}

\caption{\label{fig:Results-Stereo-Core-height}Stereo reconstruction of height
of shower core, i.e., cloud height in case of reflection off opaque
cloud. The cloud height is underestimated by approximately 0.5~km
but the resolution is better than 0.5 km.}
\end{figure}

In case of stereo geometry, the last observed point along the shower
track (pointing direction) can be converted to a position in space
and therefore a height above the surface. This point can be interpreted
as the surface height or cloud top height. Cloud presence can be identified
by comparing this height with the known surface elevation. Figure
\ref{fig:Results-Stereo-Core-height} shows results from a test study.
The cloud height is underestimated by approximately 0.5~km but the
resolution is better than 0.5 km. In the figure, the error bars indicate
the spread in the calculated heights and not the error on the mean. 

\section*{Acknowledgments}

This research was supported by NSF grants number PHY-9974537 and PHY-0140688,
also by NASA grant number NAG53902. We would like to thank Prof. Ellsworth
and the Physics Dept. at George Mason University for the use of their
facilities.

\appendix

\section{\label{sec:Clouds-Simulation}Cloud Simulation}

\subsection{\label{sub:Clouds}Introduction}

The input parameters for the clouds model are: the cloud base height,
$h_{base}$, the top height, $h_{top}$, and the vertical optical
depth, $\tau$. The clouds density is uniform and falls to zero at
the boundaries:

\[
\rho_{c}=\left\{ \begin{array}{cc}
1 & h_{base}<h<h_{top}\\
0 & otherwise\end{array}\right.\]
The vertical extent of the cloud, $\Delta z=h_{top}-h_{base}$, and
the optical depth, $\tau$, determine the scattering coefficient through
the relation $\tau=\beta\Delta z$. With $\beta$ given in 1/m. Note
that $\sigma_{e}=\beta$ since we assume no absorption. The wavelength
dependence of the cloud's optical parameters is mild for $\lambda<0.5$~$\mu$m
\cite{Liou_fig5.3} and is ignored. 

Several phase functions are relevant to a discussion of cloud scattering.
The simplest is the Henyey-Greenstein (HG) phase function \cite{HenGreen},
given in eq. \ref{eq:HG}. It is often used in radiative transfer
calculations as an analytic approximation to actual phase functions
which may display complicated structures, see \cite{Boucher_98} and
references therein.

\begin{equation}
P_{HG}(\cos(\theta_{s});g)=\frac{(1-g^{2})}{\left[1+g^{2}-2g\cos(\theta_{s})\right]^{3/2}}\label{eq:HG}\end{equation}
 Note that the parameter $g$ in the HG function is equal to the \emph{asymmetry}
\emph{parameter} defined by:

\[
g=\frac{1}{2}\int_{-1}^{1}P(\cos(\theta_{s}))\cos(\theta_{s})d\cos(\theta_{s})\]
 where $\theta_{s}$ is the scattering angle and $P(\cos(\theta_{s}))$
is the phase function.

\begin{figure}
\begin{center}\includegraphics[%
  scale=0.45]{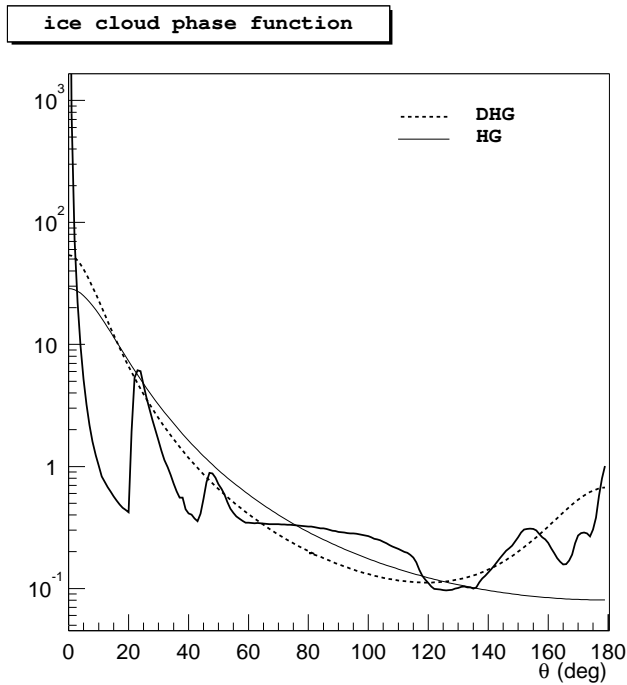}\end{center}

\caption{\label{fig:ice_phase}Ice clouds phase function (data from Liou at
0.5~$\mu$m .) A double Henyey-Greenstein approximation also shown.}
\end{figure}

One feature of realistic clouds phase functions is a backward scattering
peak. This feature is not reproduced by the HG function, however,
a \emph{double-Henyey-Greenstein} (DHG) function can provide a better
fit, see fig. \ref{fig:ice_phase}. The DHG function is defined by
\cite{White_dHG}:

\[
P_{DHG}(g)=f\times P_{HG}(g_{1})+(1-f)\times P_{HG}(g_{2})\]
 where $f\approx1$ gives the forward scattering strength, and $g_{2}$
is negative.

Liou \cite{Liou_p276} gives in tabular form the phase function for
a cirrostratus cloud model at a wavelength $\lambda=500$~nm. It
is shown in figure \ref{fig:ice_phase} along with a HG and a DHG
functions superimposed. A calculation of the asymmetry parameter for
the realistic phase function gives $g=0.753$. The same value is used
in the superimposed HG function. The DHG function parameters were
set to: $g_{1}=0.82$, $g_{2}=-0.82^{2}$, and $f=0.96$.

\subsection{\label{sub:Multiple-Scattering-in}Multiple Scattering in Clouds }

\begin{figure}
\begin{center}\includegraphics[%
  scale=0.5]{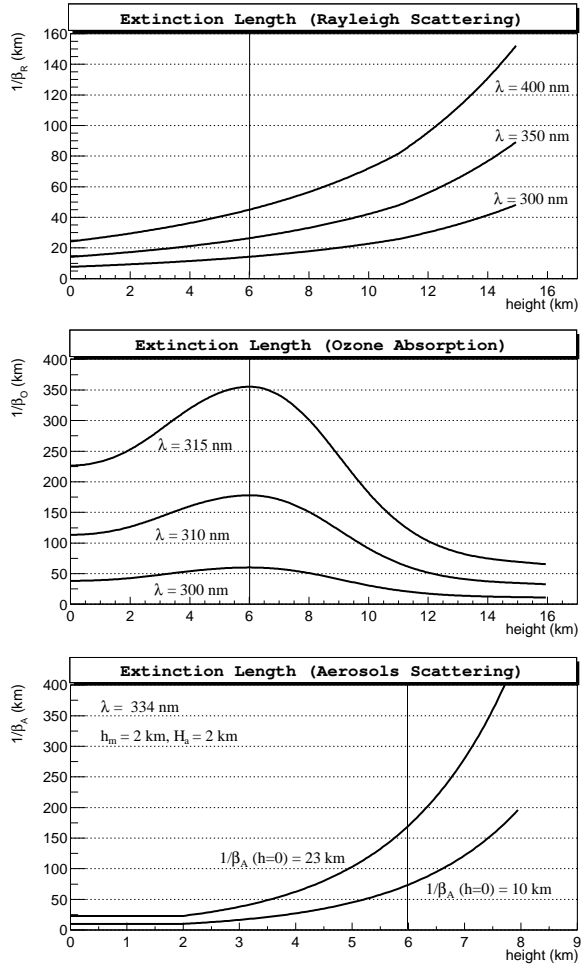}\end{center}

\caption{\label{fig:ext_len}extinction length due to Rayleigh scattering,
ozone absorption, and aerosols scattering as a function of altitude
above sea level. U.S. 1976 standard atmosphere used for density profile
for Rayleigh scattering. }
\end{figure}

A proper treatment of multiple scattering in clouds has to take into
account the scattering and absorption by other atmospheric constituents.
However, by considering the relative strength of the relevant processes,
we can show that under certain conditions it is safe to ignore some
of them. Thin cirrus clouds occur in the atmosphere at altitudes greater
than 6~km and could be as high as 15~km. The extinction length varies
with the clouds ice water content (determined in part by the altitude)
and takes values on the order of a few kilometers. For comparison,
extinction lengths due to molecular scattering, ozone absorption,
and aerosols scattering are shown in figure \ref{fig:ext_len}.

Ozone absorption is negligible for wavelengths greater than $\lambda=320$~nm,
and as can be seen from the figure, it can be safely dropped from
the multiple scattering calculation for $\lambda>310$~nm. Aerosols
density and extinction length are variable and whether or not they
can be neglected depends on the local conditions. Figure \ref{fig:ext_len}
shows two examples, $L_{a}=10$~km, which represents a hazy atmosphere,
and $L_{a}=23$~km, corresponding to an average atmosphere. Even
in hazy conditions, the attenuation length due to aerosols is large
for altitudes greater than 6~km because of the small scale height
of the aerosols density distribution. For an average atmosphere, the
aerosols extinction length is almost 20 times as large as the typical
cloud extinction length, and can be safely neglected. Rayleigh scattering
has a very strong wavelength dependence and can not be ignored at
wavelengths close to $\lambda=300$~nm. For larger $\lambda$, and
at altitudes greater than 6~km, the Rayleigh scattering length is
greater than 20~km. 

\subsubsection{An Isotropic Source}

In this section we develop expressions for the direct transmission
and transmission due to first, and second order scattering of light
from a point source inside a cloud to a detector. First we define
some notation. The point source is located at position $\vec{r}_{0}$,
the detector (mirror) at $\vec{r}_{m}$. The points $\vec{r}_{1}$,
$\vec{r}_{2}$ are the locations inside the cloud of first and second
photon scattering respectively. The distance between any of the above
points is denoted by $l_{ab}$ where $ab$ take the values of the
subscripts of the respective points. In the case of the detector,
we denote by $l_{am}$ the distance between $\vec{r}_{a}$ and $\vec{r}_{m}$
which lies \emph{inside} the cloud, i.e. the path-length inside the
cloud. the actual distance, $|\vec{r}_{m}-\vec{r}_{a}|$, is denoted
by $R_{am}$. Direction and solid angle are denoted by $\Omega$.
For example, $\Omega_{01}$ is the direction defined by the unit vector
$(\vec{r}_{1}-\vec{r}_{0})/\left|\vec{r}_{1}-\vec{r}_{0}\right|$.
The detector effective aperture is denoted by $A$, and the projected
area of the detector with respect to $\vec{r}_{0}$ by $A_{\bot}^{(0)}$.
The optical path length due to scattering by non-cloud particles (air,
aerosols, or ozone absorption) is denoted by $\tau^{(nc)}$ with subscripts
to identify the path. Non-clouds scattering and absorption coefficients
are height dependent and so an integration over the path is required.

Let the point source emit isotropically $N_{s}$ photons, then the
number of directly transmitted photons reaching the detector is given
by:

\[
N_{0}=\frac{N_{s}A_{\bot}^{(0)}}{4\pi R_{0m}^{2}}e^{-\beta_{c}l_{0m}}e^{-\tau_{0m}^{(nc)}}\]

First order scattering involves direct transmission to a point $\vec{r}_{1}$
then scattering in a volume element $dV_{1}$ into a solid angle $d\Omega$
at $\Omega$: 

\[
\frac{dN_{1}}{d\Omega}(\vec{r}_{1},\Omega)=\left(\frac{N_{s}\beta_{c}dV_{1}}{4\pi l_{01}^{2}}P(\Omega,\Omega_{01})e^{-\beta_{c}l_{01}}e^{-\tau_{01}^{(nc)}}\right)\]
 where $P(\Omega,\Omega_{01})$ is the phase function for scattering
by the cloud.

The number of photons received by the detector due to first order
scattering taking attenuation along the path from $\vec{r}_{1}$ to
the detector into account, is given by the integral over the cloud
of:

\[
dN_{1}=\frac{dN_{1}}{d\Omega}(\vec{r}_{1},\Omega_{1m})\times\left(\frac{A_{\bot}^{(1)}}{R_{1m}^{2}}e^{-\beta_{c}l_{1m}}e^{-\tau_{1m}^{(nc)}}\right)\]

The expression for second order scattering is similar to first order,
however, instead of a point source at $\vec{r}_{0}$ we now have an
integral over $\vec{r}_{1}$:

\begin{eqnarray*}
\frac{dN_{2}}{d\Omega}(\vec{r}_{2},\Omega) & = & \beta_{c}dV_{2}\int\frac{dN_{1}}{d\Omega}(\vec{r}_{1},\Omega_{12})\\
 & \times & \frac{1}{l_{12}^{2}}P(\Omega,\Omega_{12})e^{-\beta_{c}l_{12}}e^{-\tau_{12}^{(nc)}}\end{eqnarray*}

The number of photons received by the detector is given by the integral
of:

\[
dN_{2}=\frac{dN_{2}}{d\Omega}(\vec{r}_{2},\Omega_{2m})\times\left(\frac{A_{\bot}^{(2)}}{R_{2m}^{2}}e^{-\beta_{c}l_{2m}}e^{-\tau_{2m}^{(nc)}}\right)\]

The evaluation of the above expressions and their integrals is done
numerically, once a point source, a detector, and the cloud/atmosphere
are specified.

\subsubsection{\label{sub:cloud_beam_scattering}Scattering out of a beam}

Instead of an isotropic point source we now consider a light beam
propagating in the direction, $\Omega_{0}$. Here we assume that $\Omega_{0}$
\emph{does not} point toward the detector, i.e. the detector receives
no direct light from the beam. Inside the cloud, light scattered out
of the beam at a point $\vec{r}_{0}$ along the beam propagation path
and into the detector is given by:

\[
N_{1}=N_{s}P(\Omega_{0m},\Omega_{0})\times\left(\frac{A_{\bot}^{(0)}}{R_{0m}^{2}}e^{-\beta_{c}l_{0m}}e^{-\tau_{0m}^{(nc)}}\right)\]
 where $N_{s}$ is the number of photons scattered out of the beam
at the point $\vec{r}_{0}$.

Second order scattering inside the cloud is treated as follows: First
consider light scattered out of the beam at $\vec{r}_{0}$ in some
direction $\Omega_{01}$. At a point $\vec{r}_{1}$ along this direction,
the irradiance is given by:

\[
\frac{dN_{1}}{d\Omega_{01}}(\vec{r}_{1})=N_{s}P(\Omega_{01},\Omega_{0})e^{-\beta_{c}l_{01}}e^{-\tau_{01}^{(nc)}}\]
 Next, the scattering in a volume element $dV_{1}$ at $\vec{r}_{1}$
into an arbitrary direction $\Omega$ is given by:

\[
\frac{dN_{2}}{d\Omega}(\vec{r}_{1},\Omega)=\beta_{c}dV_{1}\left[\frac{dN_{1}}{d\Omega_{01}}(\vec{r}_{1})\right]P(\Omega,\Omega_{01})\]
 Finally the contribution to the detector signal from the point $\vec{r}_{1}$
is:

\[
dN_{2}=\frac{dN_{2}}{d\Omega}(\vec{r}_{1},\Omega_{1m})\times\left(\frac{A_{\bot}^{(1)}}{R_{1m}^{2}}e^{-\beta_{c}l_{1m}}e^{-\tau_{1m}^{(nc)}}\right)\]
 The integral over the cloud volume of $dN_{2}$ gives the total contribution
due to second order scattering of the beam photons inside the cloud.

\subsubsection{Approximations}

The OWL detector is located at an altitude of 800 km, this along with
the fact that the scattering length inside the cloud is on the order
of a few km's, implies that: $R_{1m}\approx R_{0m}$ and $A_{\bot}^{(1)}\approx A_{\bot}^{(0)}$.These
variables can then be taken out of the integrals and replaced by the
approximate values.

The optical path length due to non-cloud scattering between two points,
$\tau^{(nc)}$, requires an integration over the path joining the
two points. A significant reduction in computation time can be achieved
if appropriate approximations are used to replace these expressions
which appear in the integrals by average values which can be taken
out of the integrals. In the case of first order scattering we have:
$\tau_{01}^{(nc)}$ and $\tau_{1m}^{(nc)}$. Given the strong forward
peak of the scattering function we can see that the largest contribution
to the integral comes from points close to the line joining the source
and detector. This allows an approximation: $\bar{s}_{1}\approx\int dss\exp(-\beta_{c}s)/\int ds\exp(-\beta_{c}s)$
with the integration along the line segment from the source to the
detector which is contained within the cloud. A vector position $<\vec{r}_{1}>$
can be defined using $\bar{s}_{1}$. Now $\tau_{1m}^{(nc)}$ will
be replaced by an average value and taken out of the integral.

From the discussion at the beginning of this section we see that aerosols
and ozone may be ignored in the volume of the cloud. Hence, $\tau_{01}^{(nc)}\approx\tau_{01}^{(R)}$,
the optical depth due to Rayleigh scattering. The latter is given
by $\int ds\beta_{R}(h)$ where $ds$ is along the line joining $\vec{r}_{0}$
and $\vec{r}_{1}$ and $h$ is the altitude along this line. In most
cases of interest, the integral can be approximated by $|\vec{r}_{1}-\vec{r}_{0}|\times\beta_{R}(\bar{h})$,
where $\bar{h}$ is the height of the midpoint between the two positions.
This is due to the fact that the cloud thickness is on the order of
one to a few km, less than the atmosphere scale height of $\sim7$~km
so $\beta_{R}$ does not change much along the integration path. This
approximation was verified to be accurate to within 1-5\% for a large
number of test cases. 

The strong wavelength dependence of Rayleigh scattering implies that
the calculation should be repeated for each wavelength of interest.
However, after considering a number of cloud configuration we saw
that the result changes by less than 20\% for wavelengths in the range
of 337~nm, and 391~nm. We concluded that it would be a reasonable
approximation to perform the calculation at a wavelength of $\lambda=357$~nm,
and use the result as an average to be taken out of the sum over wavelengths.

Finally, after the calculation outlines in section \ref{sub:cloud_beam_scattering}
was implemented for the shower's Cerenkov beam, it became apparent
that a simple alternative calculation which accounts for most of the
additional signal received by the detector can be used instead. The
strong forward peak of the cloud phase function results in that more
than 52\% of the photons scattered out of a beam are scattered forward
in a cone of half-angle of 2$^{o}$. By not subtracting these photons
from the beam, we in effect calculate the second order scattering
of these photons at a later stage along the beam propagation when
we evaluate the first order scattering from the beam at the later
stage. The other 48\% photons neglected in this approximation will
have a lesser effect on the detector signal once one considers finite
time window for the detector pixels.

\subsection{\label{sub:The-case-tau-gt3}Simple Cloud Monte Carlo}

A simple method which works well and serves our needs is the Monte
Carlo (MC) method. MC calculations are valid for clouds of all optical
depths, however we only employ them for optically thick clouds. Currently
our implementation only allows for clouds scattering but it can be
easily extended to include Rayleigh and aerosols scattering. The calculation
involves the following steps;

\begin{enumerate}
\item Select photon initial position inside the cloud or at a cloud boundary.
Also, select the photon direction and a time offset relative to some
$t_{0}$. The photon direction can be random, for isotropic distribution,
or fixed in case of a beam.
\item Propagate the photon by a random step (distance) chosen from an exponential
distribution: $\exp(-\beta_{c}l)$, with $\beta_{c}$ the extinction
coefficient of the cloud
\item Check if new photon position is inside the cloud. If not then done,
if it is then continue.
\item Select a random scattering angle drawn from a distribution which follows
the clouds phase function. Select a uniform azimuthal angle. Set new
photon direction.
\item Goto step 2
\end{enumerate}
A simple {}``cloud MC'' was developed around this algorithm to calculate
the beam reflection from the top of a cloud. For a cloud with given
cloud parameters a large number of photons impinging on the cloud
top at a fixed angle (representing the beam's zenith angle) is followed
through the cloud. As the photons emerge from the cloud, either the
cloud top (reflected) or cloud bottom (transmitted) they are added
to a set of histograms which record the distributions of the locations
and time delays of the photons. At the end of the run the histograms
are saved to file and can be later used by the detector MC.

\bibliographystyle{unsrt}
\bibliography{/home/tareq/bib/astro,/home/tareq/bib/atmos,/home/tareq/bib/eas,/home/tareq/bib/misc,/home/tareq/bib/owl,/home/tareq/bib/tareq}

\end{document}